\documentclass[paper,a4paper,11pt,twoside]{JHEP3}
\usepackage{psfrag}
\usepackage{epsfig}
\usepackage{amsmath}
\usepackage{multirow}
\usepackage{amssymb}
\usepackage{bbm}
\usepackage{cite}
\newcommand{\be}{\begin{equation}}
\newcommand{\ee}{\end{equation}}
\newcommand{\ba}{\begin{array}}
\newcommand{\ea}{\end{array}}
\newcommand{\bi}{\begin{itemize}}
\newcommand{\ei}{\end{itemize}}
\newcommand{\oh}{{\frac 12}}

\newcommand{\fa}{f_{\rm A}}
\newcommand{\fp}{f_{\rm P}}
\newcommand{\fv}{f_{\rm V}}

\newcommand{\ca}{c_{\rm A}}
\newcommand{\cv}{c_{\rm V}}
\newcommand{\za}{Z_{\rm A}}
\newcommand{\zp}{Z_{\rm P}}
\newcommand{\zv}{Z_{\rm V}}
\newcommand{\zM}{Z_{\rm M}}
\newcommand{\zm}{Z_{\rm m}}
\newcommand{\bA}{b_{\rm A}}
\newcommand{\bp}{b_{\rm P}}
\newcommand{\bv}{b_{\rm V}}
\newcommand{\bm}{b_{\rm m}}

\newcommand{\Ds}{\dsub}

\newcommand{\ds}{{\rm D_s}}
\newcommand{\MSbar}{\overline{{\rm MS}}}
\newcommand{\dsubstar}{{\rm{D}_s^\ast}}
\newcommand{\fDs}{F_{\rm D_s}}
\newcommand{\fDsstar}{F_{\rm D_s^\ast}}
\newcommand{\mds}{{m_{\rm{D}_s}}}
\newcommand{\mdsstar}{{m_{\rm{D}_s^\ast}}}
\newcommand{\dsub}{{\rm{D}_s}}

\newcommand{\Exp}{{\rm e}}
\title{Lattice cutoff effects for $\fDs$ with improved Wilson fermions 
-- a final lesson from the quenched case}

\author{%
\begin{flushleft}
\vspace{-0.5cm}
\vbox{
\epsfxsize=2.5 true cm
\epsfbox{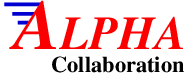}}
\end{flushleft}
}

\author{%
Jochen Heitger\\
Westf\"alische Wilhelms-Universit\"at M\"unster,
Institut f\"ur Theoretische Physik,\\
Wilhelm-Klemm-Strasse~9, D-48149 M\"unster, Germany\\
E-mail: \email{heitger@uni-muenster.de}
}%
\author{%
Andreas J\"uttner\\
Johannes Gutenberg Universit\"at Mainz,
Institut f\"ur Kernphysik,\\
Johann-Joachim-Becher Weg 45, D-55099 Mainz, Germany\\
E-mail: \email{juettner@kph.uni-mainz.de}}

\preprint{%
MS-TP-08-34\\
MKPH-T-08-19\\
\today
}

\abstract{%
In view of the 
recent excitement about a tension between determinations of 
$\fDs$ from experiment and from simulations of lattice QCD with 
dynamical quarks, we try to clear up the picture of lattice determinations
in the continuum limit of the 
quenched approximation. For ${\rm O}(a)$ improved Wilson quarks 
we see linear scaling in the squared lattice spacing $a^2$ only for 
$a\lesssim0.08\,$fm. For coarser lattices we observe significant 
contaminations from higher order cutoff effects.
As an aside we also study the scaling of the charm quark mass and the ratio
of the vector to the pseudo-scalar decay constant and the spin-splitting.
}

\begin{document}
%
%%%%%%%%%%%%%%%%%%%%%%%%%%%%%%%%%%%%%%%%%%%%%%%%%%%%%%%%%%%%%%%%%%%%%%%%%%%%%%
\section{Introduction}
%%%%%%%%%%%%%%%%%%%%%%%%%%%%%%%%%%%%%%%%%%%%%%%%%%%%%%%%%%%%%%%%%%%%%%%%%%%%%%
%
Since the time when simulations of lattice QCD with dynamical fermions 
became feasible, the phenomenology of the $\ds$-meson has been considered 
as a field where lattice QCD could provide benchmark predictions that would
be confronted with increasingly precise measurements from experiments 
like Belle, BaBar and CLEO-c.
Preparatory tests of the techniques in quenched lattice QCD 
\cite{Juttner:2003ns,AliKhan:2007tm}
indicated that a precision of only a few percent for observables
like $\fDs$ is possible when simulating the full theory, while
keeping all sources of systematic uncertainties under control\footnote{
The systematic effect due to the chiral extrapolation of the sea quark masses
cannot be assessed in the quenched theory.}.

Only recently, precise determinations of $\fDs$ appeared in dynamical
simulations with $N_{\rm f}=2$ and 2+1 flavours of sea quarks
\cite{Aubin:2005ar,Follana:2007uv,ETMCLAT08,vonHippel:2008pc}. 
The central values, although partly preliminary 
\cite{ETMCLAT08,vonHippel:2008pc}, turned out to be about 10\% larger 
than in the quenched case. This was not surprising given that the quenching
error was always estimated to lie in the range of 10--20\%. 
Surprising, however, turned out to be a recent comparison with a 
compilation of experimental results for $\fDs$
\cite{Rosner:2008yu} 
by CLEO-c \cite{Chadha:1997zh} and Belle \cite{:2007zm,Artuso:2007zg}. 
A tension between the experimental value and 
lattice results triggered speculations about possible signs for physics
beyond the Standard Model \cite{Dobrescu:2008er}.

In this work we want to emphasize that
the lattice spacing dependence must be mapped out over a large range in order 
to unambiguously isolate the leading lattice artefacts -- 
this is all the more important in the heavy quark sector.
In particular the approach to the continuum limit of the leptonic 
decay constant of the $\dsub$-meson in the quenched approximation,
as recently summarized in \cite{AliKhan:2007tm,DellaMorte:2007ny}, 
did not exhibit proper scaling, and we try to clarify this issue here. 
It appears that simulations with lattice spacings larger than
$a\approx 0.08\,$fm can suffer from ambiguities in the ${\rm O}(a)$ 
improvement of the quark bilinear currents, which significantly affects
the extrapolation of the data to the continuum limit. 
It is indispensable to go to smaller lattice spacings in order to assure 
scaling. 
These findings are extremely important for assessing current error 
estimates in lattice studies of $\fDs$ in the unquenched case as they 
have been reported 
by ALPHA, ETMC, HPQCD and MILC 
\cite{vonHippel:2008pc,ETMCLAT08,Follana:2007uv,Aubin:2005ar}. 

To this end we extend the ALPHA Collaboration's computation of $\fDs$ in 
the quenched approximation reported in \cite{Juttner:2003ns} for lattice 
spacings in the range $a\approx0.09-0.05\,$fm by simulations of an additional 
lattice spacing $a\approx0.03\,$fm \cite{Juttner:2004tb}. 
Moreover, we increased the statistics with respect to \cite{Juttner:2003ns} 
and \cite{Juttner:2004tb} significantly, which allows us to
present a comprehensive picture of the lattice spacing dependence of the 
decay constant $\fDs$ in quenched lattice QCD with 
non-perturbatively improved Wilson quarks. In addition, we also present
scaling studies for the renormalization group invariant 
charm quark mass $M_{\rm c}$ \cite{Rolf:2002gu,Juttner:2004tb}, 
the mass splitting  $m_{\dsubstar}-m_{\dsub}$ and of the ratio of 
the vector to pseudo-scalar meson
decay constants $\fDsstar/\fDs$ \cite{Juttner:2004tb}.
We start by briefly explaining the setup of our large-volume simulations of 
the quenched QCD Schr\"odinger functional \cite{Luscher:1992an,Sint:1994un} 
and then discuss the simulation parameters. This is succeeded by a 
presentation of our results, before ending with a discussion and our 
conclusion.
%
%%%%%%%%%%%%%%%%%%%%%%%%%%%%%%%%%%%%%%%%%%%%%%%%%%%%%%%%%%%%%%%%%%%%%%%%%%%%%%
\section{Details of the simulation}
%%%%%%%%%%%%%%%%%%%%%%%%%%%%%%%%%%%%%%%%%%%%%%%%%%%%%%%%%%%%%%%%%%%%%%%%%%%%%%
%
In the following we introduce the set of Schr\"odinger functional
correlation functions, from which we determine
$\mds$, $\mdsstar$, $\fDs$, $\fDsstar$ and the mass of the charm quark.
We compute the correlation functions directly for 
all required mass-degenerate and non-degenerate combinations built from
quark propagators at the physical strange quark mass and at two values of 
the charm quark mass close to the physical one. 
This allows for an interpolation of the results 
to the exact physical point, which we define through 
the experimentally determined mass of the $\Ds$-meson as input.
This computation is repeated for five different lattice spacings in the range 
of $a\approx 0.1 - 0.0\,3\,$fm 
in order to investigate the scaling behaviour of our observables and
identify a  
range in the lattice spacing where observables scale linearly in~$a^2$.
%%%%%%%%%%%%%%%%%%%%%%%%%%%%%%%%%%%%%%%%%%%%%%%%%%%%%%%%%%%%%%%%%%%%%%%%%%%%%%
\subsection{Correlation functions}
%%%%%%%%%%%%%%%%%%%%%%%%%%%%%%%%%%%%%%%%%%%%%%%%%%%%%%%%%%%%%%%%%%%%%%%%%%%%%%
%
For this study the relevant Schr\"odinger functional heavy-light 2-point 
correlation functions are
\begin{equation}
 \begin{array}{l}
  \fa(x_0)=-\frac{1}{2}\langle A_0(x){\cal O} \rangle\,,\ \
  \fp(x_0) = -\frac{1}{2}\langle P(x){\cal O} \rangle\,,\ \
  f_1 = - \frac{1}{2L^6} \langle {\cal O}' {\cal O}\rangle\,,\label{eqn:cfs}
	\\[2mm]
  \fv(x_0) = -\frac{1}{6}\sum\limits_i
		\langle V_i(x){\cal O}_i \rangle\,,\ \
  f_{{\rm T},\mu}(x_0) = -\frac{1}{6}\sum\limits_i\langle T_{\mu i}{\cal O}_i(x)
                      \rangle\,,\ \
  k_1 = - \frac{1}{6L^6}\sum\limits_i \langle{\cal O}^\prime_i {\cal O}_i
        \rangle\,,
 \end{array}
\end{equation}
where 
$A_{\mu}(x) = \overline{\psi}_{\rm s}(x)\gamma_{\mu}\gamma_5\psi_{\rm h}(x)$ 
is the axial-vector current,
$P(x)= \overline{\psi}_{\rm s}(x)\gamma_5\psi_{\rm h}(x)$ 
is the pseudo-scalar density,
and $V_{\mu}(x) = \overline{\psi}_{\rm s}(x)\gamma_{\mu}\psi_{\rm h}(x)$ and
$T_{\mu\nu}(x) = \overline{\psi}_{\rm s}(x)\sigma_{\mu\nu}\psi_{\rm h}(x)$ are
the vector and tensor currents. 
In each case we indicate the respective quark flavour
with a subscript ``${\rm s}$'' for light ($=$ strange) and ``${\rm h}$'' for 
heavy quarks with masses close to the one of the charm quark. 
The quark bilinears
\begin{eqnarray}
 {\cal O} & = & \frac{a^6}{L^3}\sum_{{\bf y}, {\bf z}}
 \overline{\zeta}_{\rm h}({\bf y})\gamma_5\zeta_{\rm s}({\bf z})\,,\quad 
 {\cal O}' = \frac{a^6}{L^3}\sum_{{\bf y}, {\bf z}}
 \overline{\zeta}'_{\rm s}({\bf y})\gamma_5\zeta'_{\rm h}({\bf z})\,,\\
 {\cal O}_i & = & \frac{a^6}{L^3}\sum_{{\bf y}, {\bf z}}
 \overline{\zeta}_{\rm h}({\bf y})\gamma_i \zeta_{\rm s}({\bf z})\,,\quad 
 {\cal O}^\prime_i = \frac{a^6}{L^3}\sum_{{\bf y}, {\bf z}}
 \overline{\zeta}'_{\rm s}({\bf y})\gamma_i\zeta'_{\rm h}({\bf z})\,,
\end{eqnarray}
are the meson wall sources at the $x_0=0$ and the $x_0=T$ boundary
timeslices, respectively.  For more details  on 
Schr\"odinger functional correlation functions
and for unexplained notation we refer the reader to, e.g.,
refs.~\cite{Luscher:1996sc,Guagnelli:1999zf}. 
We employ the improved axial-vector and vector currents 
\begin{equation}
   A_{0}^{\rm I}(x) = A_{0}(x) + \ca a \tilde{\partial}_{0} P(x)
		\quad{\rm and}\quad
   V_{i}^{\rm I}(x) = V_{i}(x) + \cv a \tilde{\partial}_{0} T_{0 i}(x)\,,
\end{equation}
where $\tilde{\partial}_{0}$ is the time component of the 
symmetrized next-neighbour lattice derivative.
The improvement coefficients $\ca(g_0)$ and $\cv(g_0)$ 
for the quenched theory
have been determined 
non-perturbatively in \cite{Guagnelli:1998db} and 
\cite{Luscher:1997ug,Bhattacharya:2001ks}, respectively\footnote{In 
the case of $\cv(g_0)$, 
the parameterization derived in \cite{Harada:2002jh} has been used here.}.
On the lattice, the axial-vector current and the vector current
receive a multiplicative (scale independent) renormalization and we write
\begin{eqnarray}
\label{eq:AR}
  (A_{\rm R})_{0} = \za \left[1 + \bA (am_{{\rm q},{\rm h}} 
  + am_{{\rm q},{\rm s}})/2 \right]A_{0}^{\rm I} + {\rm O}(a^2)\,, \\
\label{eq:VR}
  (V_{\rm R})_{i} = \zv \left[1 + \bv (am_{{\rm q},{\rm h}} 
  + am_{{\rm q},{\rm s}})/2 \right]V_{i}^{\rm I} + {\rm O}(a^2)\,. 
\end{eqnarray}
The corresponding renormalization constants 
for the quenched theory
$\za(g_0)$ and $\zv(g_0)$ have  
been determined non-pertur\-bative\-ly in 
\cite{Luscher:1997jn}.
${\rm O}(a)$ artifacts that are proportional to the bare
subtracted quark masses 
$am_{{\rm q},i}=
\frac{1}{2}\big(\frac{1}{\kappa_i}-\frac{1}{\kappa_{\rm crit}}\big)$
for $i={\rm s},{\rm h}$,
where $\kappa_{\rm crit}$ is the critical hopping parameter, 
are canceled by terms proportional to the improvement coefficients 
$\bA(g_0)$ and $\bv(g_0)$; 
they have been calculated in 1-loop perturbation theory \cite{Luscher:1997ug}.
%
%%%%%%%%%%%%%%%%%%%%%%%%%%%%%%%%%%%%%%%%%%%%%%%%%%%%%%%%%%%%%%%%%%%%%%%%%%%%%%
\subsection{Observables}
%%%%%%%%%%%%%%%%%%%%%%%%%%%%%%%%%%%%%%%%%%%%%%%%%%%%%%%%%%%%%%%%%%%%%%%%%%%%%%
%
We define the pseudo-scalar effective mass via
\begin{equation}
\label{eq:meff}
  am_{\text{PS}}(x_0+a/2) = \log\frac{\fa^{\rm I}(x_0)}{\fa^{\rm I}(x_0+a)}\,, 
\end{equation}
and the expression for the local pseudo-scalar meson decay constant is 
given by \cite{Guagnelli:1999zf}
\begin{eqnarray}
\label{eq:fDs}
a{F_{\rm PS}}(x_0) 
&=& 
-2 \za \left[1+\bA (am_{{\rm q},{\rm h}}+am_{{\rm q},{\rm s}})/2 \right]
\frac{\fa^{\rm I}(x_0)}{\sqrt{f_1}} \,
(m_{\rm PS}L^3)^{-1/2} \,\Exp^{\,(x_0-T/2) m_{\rm PS}} \nonumber\\
& &
\times \left\lbrace 1 - \eta_{\rm A}^{\rm PS} \Exp^{\,-x_0\Delta} - \eta_{\rm A}^0
\Exp^{\,-(T-x_0) m_{\text{G}}}\right\rbrace + {\rm O}(a^2)\,,
\end{eqnarray}
and an analogous expression holds for the mass and the decay constant in 
the vector channel.
In (\ref{eq:fDs}) 
the factor $(m_{\rm PS}L^3)^{-1/2}$ takes into account the
normalization of one-particle states, and $f_1^{-1/2}$
cancels out the dependence on the meson sources. 
Because of its exponential decay, the correlation function $\fa^{\rm I}$ is 
dominated by the ground state in the pseudo-scalar channel for large time 
separations from the boundaries (i.e.~$x_0$ and $T-x_0$).
Hence, eqs.~(\ref{eq:meff}) and (\ref{eq:fDs}) are expected to exhibit a 
plateau at intermediate times, when the contribution 
$\eta_{\rm A}^{\rm PS}\Exp^{\,-x_0\Delta}$ of the first excited state and the 
contribution $\eta_{\rm A}^0 \Exp^{\,-(T-x_0)m_{\text{G}}}$ from the $O^{++}$ 
glueball both are 
small and well below the statistical errors,
such that final estimates for masses and decay constants are obtained by
fits to a constant within the plateau regions of the local quantities 
(\ref{eq:meff}) and (\ref{eq:fDs}), as explained in~\cite{Guagnelli:1999zf}.

Concerning the quark masses, we employ the continuum PCAC relation, which
in the situation of QCD with non-degenerate quarks at hand can be written 
as 
\be
\partial_\mu A_\mu(x)=2m_{\rm hs}P(x)=(m_{\rm hh}+m_{\rm ss})P(x)
\ee
and on the lattice translates into a definition of the bare current quark
mass in terms of the heavy-light correlation functions (\ref{eqn:cfs})
\cite{Luscher:1996sc}, 
\be\label{barePCACmass}
am_{\rm hs}(x_0) = 
\frac a2 \left[\oh(\partial^\ast_0+\partial_0)\fa(x_0)
+ \ca a \partial_0^\ast\partial_0 \fp(x_0)\right]/\fp(x_0)\,,
\ee
$\partial_0$ ($\partial^\ast_0$) being the forward (backward) lattice 
derivative in the time direction.
In this notation, a single quark mass $m_{ii}$, $i={\rm s},{\rm h}$, 
is obtained if the quark flavours are mass degenerate.
Again it is understood that the bare heavy-light current quark masses 
entering the formulae below are computed as timeslice averages within a 
central flat region of the associated local masses, 
eq.~(\ref{barePCACmass}).

After multiplicative renormalization of the improved axial current and
the pseudo-scalar density we obtain the identity
\be\ba{rcl}\label{PCACmass}
a(m_{{\rm R},{\rm h}}+m_{{\rm R},{\rm s}})
&=&
\displaystyle 
2\,\frac{\za\left[1+\bA\oh(am_{{\rm q},{\rm h}}+am_{{\rm q},{\rm s}})\right]}
{\zp\left[1+\bp\oh(am_{{\rm q},{\rm h}}+am_{{\rm q},{\rm s}})\right]}\,
am_{\rm hs}+{\rm O}(a^2)\\\\
&=&
2\za\zp^{-1}\left[
1+(\bA-\bp)\oh(am_{{\rm q},{\rm h}}+am_{{\rm q},{\rm s}})
\right]am_{\rm hs}+{\rm O}(a^2)
\ea
\ee
for the sum of the renormalized PCAC heavy and strange (valence) quark masses
in the O($a$) improved theory, where $(\bA-\bp)(g_0)$ has been determined 
non-pertur\-bative\-ly in \cite{Guagnelli:2000jw}.
The scale dependent (and, in quark mass independent renormalization schemes, 
also flavour independent) renormalization factor $\zp=\zp(g_0,L/a)$ encodes 
the 
scale dependence of the renormalized quark masses and is 
available from \cite{Capitani:1998mq} in the Schr\"odinger functional scheme 
for $N_{\rm f}=0$ and the range of bare couplings relevant here.  
Eventually, using the flavour independent ratio $M/m_{{\rm R},i}(\mu)$, 
$\mu=1/L$, in the continuum limit, we combine this with $\za/\zp$ into the 
total renormalization factor $\zM(g_0)$ \cite{Capitani:1998mq} to express 
our results directly in terms of the scale and scheme independent 
renormalization group invariant (RGI) quark mass $M$.
The RGI mass is very convenient 
since it
may then be straightforwardly converted to any other 
renormalization scheme, such as $\MSbar$ at some desired scale, by means of 
continuum perturbation theory.

From the various quark mass definitions for non-degenerate and degenerate 
quarks we construct the following expressions for the O($a$) improved RGI 
heavy (i.e.~charm) quark mass.
Once we compute it via the current quark mass from the heavy-strange current 
with the strange current quark mass subtracted,
\be\ba{rcl}
aM_{\rm h}^{({\rm hs})} 
&=& 
\zM\,\Bigl\{2 m_{\rm hs}
\left[ 1+(\bA-\bp){1\over 2} (am_{\rm q,h}+am_{\rm q,s})\right]\\
& &
-\,am_{\rm ss}\left[1+(\bA-\bp)am_{\rm q,s}\right]\Bigr\}\,,\label{eqn:Mhs}
\ea\ee
and once from the case of degenerate quarks via the current quark mass 
from the heavy-heavy current,
\be
aM_{\rm h}^{({\rm hh})} =  
\zM \,am_{\rm hh} \left[1+(\bA-\bp)am_{{\rm q},{\rm h}}\right]\,.
\label{eqn:Mhh}
\ee
A third definition of the RGI charm quark mass is given directly in terms of 
the bare subtracted heavy quark mass through the relation 
\be
aM_{\rm h}^{({\rm h})} = 
\zM Z \,am_{{\rm q},{\rm h}}\left[1+\bm am_{\rm q,h}\right]\,.
\label{eqn:Mh}
\ee
We will use the quenched parameterization of $\zM(g_0)$ from 
\cite{Juttner:2004tb}, which extends the one based on 
\cite{Capitani:1998mq,Guagnelli:2004za} to also include $\beta=6.7859$
(see below), and of $Z(g_0)=(\zm\zp/\za)(g_0)$ and $\bm(g_0)$ from 
\cite{Guagnelli:2000jw}.
These three definitions of the quark mass will have different cutoff effects, 
but should agree with each other once being extrapolated to the continuum 
limit. This provides a nice check of the extrapolation procedure.
%
%%%%%%%%%%%%%%%%%%%%%%%%%%%%%%%%%%%%%%%%%%%%%%%%%%%%%%%%%%%%%%%%%%%%%%%%%%%%%%
\subsection{Parameters for the scaling study}
%%%%%%%%%%%%%%%%%%%%%%%%%%%%%%%%%%%%%%%%%%%%%%%%%%%%%%%%%%%%%%%%%%%%%%%%%%%%%%
%
We have generated two ensembles of gauge field configurations 
(called A and B in the following) with standard algorithms, for five 
different lattices of approximately constant physical size $L/r_0\approx 3$ 
in spatial directions, but decreasing lattice spacing.
Here, $r_0$ is the Sommer scale \cite{Necco:2001xg} which we use in order
to estimate the lattice spacing $a$ in physical units.
Some lattice and simulation parameters are collected in table 
\ref{scalingparams}.
The numerical simulations\footnote{
All simulations were done in double precision arithmetic, except for the
subset of simulation points of data set A which were the basis of
\cite{Juttner:2003ns}. However, we did not notice any visible impact of
the precision used for the charm quark propagator computation on our 
results.}
were carried out on the APEmille and apeNEXT 
computers at DESY Zeuthen and on the IBM p690 computers of the 
HLRN \cite{HLRN:2002}. 
In the latter case we have employed an adapted and performance-improved 
version of the Schr\"odinger functional implementation
based on the MILC code \cite{MILC,Juttner:2004tb}.

\begin{table}
\begin{center}
\begin{tabular}[h]{l@{\hspace{ 2mm}}cccc|ccc|ccc}
\hline
\hline
&&&&&&&&\\[-4mm]
        &&&&&\multicolumn{3}{c|}{data set A}&\multicolumn{3}{c}{data set B}\\
	&$L/a$	&$T/a$	&$L/r_0$	&$a\,[{\rm fm}]$&$n_{\rm meas}$&$n_{\rm update}$&$n_{\rm OR}$&$n_{\rm meas}$&$n_{\rm update}$&$n_{\rm OR}$\\[.5ex]
\hline&&&&&&&&\\[-2ex]
$\beta_1=6.0$ 	&$16$&$32$&$2.98$ 	&$0.093$&380&100&8  &2100 &25 &5 \\
$\beta_2=6.1$ 	&$24$&$40$&$3.79$ 	&$0.079$&201&100&12 &1300 &25 &5 \\
$\beta_3=6.2$	&$24$&$48$&$3.26$ 	&$0.068$&251&100&12 &1300 &25 &5 \\
$\beta_4=6.45$ 	&$32$&$64$&$3.06$ 	&$0.048$&289&100&12 &1400 &40 &10\\
$\beta_5=6.7859$     &$48$&$96$	&$3.00$	&$0.031$&150&50 &24 &604  &50 &20\\	[.5ex]
\hline
\hline
\end{tabular}
\caption{Lattice geometries and simulation parameters used in the simulations 
for the scaling study for data set A and data set B.
The correlation functions were evaluated on $n_{\rm meas}$ gauge 
configurations, which were separated by $n_{\rm update}$ update iterations 
consisting of one heat-bath and $n_{\rm OR}$ over-relaxation sweeps.}
\label{scalingparams}
\end{center}
\end{table}
%
%%%%%%%%%%%%%%%%%%%%%%%%%%%%%%%%%%%%%%%%%%%%%%%%%%%%%%%%%%%%%%%%%%%%%%%%%%%%%%
\subsection{Hopping parameters}\label{Hoppsandstops}
%%%%%%%%%%%%%%%%%%%%%%%%%%%%%%%%%%%%%%%%%%%%%%%%%%%%%%%%%%%%%%%%%%%%%%%%%%%%%%
The values of the critical hopping parameter $\kappa_{{\rm crit}}$,
which enter the analysis through the quark mass dependent O($a$) terms
in eqs.~(\ref{eq:AR}) and (\ref{eq:VR}) and the subsequent expressions,
were gathered from the non-perturbative determination in quenched QCD
of \cite{Luscher:1997ug} and, where necessary, we interpolated 
$\kappa_{{\rm crit}}$ to the desired value of the lattice coupling 
\cite{Juttner:2004tb}.

$\kappa_{\rm s}$ was fixed prior to the simulations using published results 
for the PCAC relation
\be\label{M_m}
M_{\rm s}+\hat M = \zM\,{{F}_{\rm K}\over G_{\rm K}}\,m_{\rm K}^2
\ee
for ${\rm O}(a)$ improved Wilson fermions in quenched QCD 
\cite{Garden:1999fg}, where $M_{\rm s}$ is the RGI quark mass of the 
strange quark and $\hat M=\oh(M_{\rm u}+M_{\rm d})$ the average RGI light 
quark mass. 
$F_{\rm K}$ is the kaon decay constant, and $G_{\rm K}$ denotes the 
vacuum-to-K matrix element of the pseudo-scalar density.
Note that in contrast to \cite{Juttner:2003ns} we here have determined 
$\kappa_s$ for each value of $\beta$ by fixing the RGI strange quark mass to
its value found at \textit{finite lattice} spacing in \cite{Garden:1999fg}.
Therefore, $\kappa_s$ differs from the one in \cite{Juttner:2003ns}
by $O(a^3)$.
We take over the values for $\zM\,{{F}_{\rm K}\over G_{\rm K}}$ 
from \cite{Garden:1999fg} at $\beta\in\{6,6.1,6.2,6.45\}$ and extrapolate 
those linearly, in order to arrive at the estimate 
$\zM{{F}_{\rm K}\over G_{\rm K}}\,|_{\,\beta=6.7859}=0.2268(57)$.
Using (\ref{M_m}) and
the ratio $M_{\rm s}/\hat M=24.4\pm1.5$ from continuum chiral perturbation 
theory~\cite{Leutwyler:1996qg}, the values for $\kappa_{\rm s}(\beta)$ are 
then determined as the solutions of 
\be\label{RGI_curr}
r_0(M_{\rm s}+\hat M) =
\left(\frac{r_0}{a}\right) \zM Z\,am_{{\rm q},{\rm s}}
\left[1+\bm am_{{\rm q},{\rm s}}\right]
\Big(1+\frac{\hat M}{M_{\rm s}}\Big)\,,
\ee
where the r.h.s. is meant to be read as a function of $\beta=6/g_0^2$.

The choice of hopping parameters of the charm quark, $\kappa_{{\rm c}_2}$,
relies on the results for $M_{\rm c}$ at $\beta\in\{6,6.1,6.2,6.45\}$ 
in \cite{Rolf:2002gu}, which in addition were extrapolated linearly in 
$(a/r_0)^2$ to $\beta=6.7859$. 
The estimates for $\kappa_{{\rm c}_2}$ in table \ref{hoppingparams}
have then been obtained by solving a quadratic equation similar to 
(\ref{RGI_curr}). We have also generated data for the supplementary 
hopping parameters $\kappa_{{\rm c}_1}^{\rm A}$ and $\kappa_{{\rm c}_1}^{\rm B}$ 
(for data set A and data set B, respectively), which yield results close 
to $\kappa_{{\rm c}_2}$.
This enabled us to interpolate all observables to the point corresponding 
to the physical charm quark.

\begin{table}
\begin{center}
\begin{tabular}{c|ccccc}
\hline\hline\\[-4mm]
$\beta$ &6              &6.1            &6.2            &6.45           &6.7859\\
\hline\\[-5mm]
$\kappa_{\rm crit}$
         &0.135196      & 0.135496      & 0.135795      & 0.135701      & 0.135120\\
\hline\\[-5mm]
$\kappa_{\rm s}$&0.134108        &0.134548   &0.134959   &0.135124   &0.134739\\
\hline\\[-5mm]
$\kappa_{{\rm c}_1}^{\rm A}$&0.123010     &0.125870       &{0.127470}     &0.130030       &0.132440 \\
$\kappa_{{\rm c}_1}^{\rm B}$&0.123010     &0.125870       &{0.127470}     &0.130030       &0.130823 \\
\hline\\[-5mm]
$\kappa_{{\rm c}_2}$& 0.119053 &0.122490 &0.124637 &0.128131 &0.130253 \\
\hline\hline
\end{tabular}
\caption{Summary of all hopping parameters. }\label{hoppingparams}
\end{center}
\end{table}
%
%%%%%%%%%%%%%%%%%%%%%%%%%%%%%%%%%%%%%%%%%%%%%%%%%%%%%%%%%%%%%%%%%%%%%%%%%%%%%%
\section{Analysis and results}
%%%%%%%%%%%%%%%%%%%%%%%%%%%%%%%%%%%%%%%%%%%%%%%%%%%%%%%%%%%%%%%%%%%%%%%%%%%%%%
%
\begin{table}
\begin{center}
\begin{tabular}{l|cccc|cccccc}
\hline\hline&&&\\[-4mm]
&\multicolumn{4}{c|}{data set A}&\multicolumn{2}{c}{data set B}\\
$\beta$&$r_0m_{\rm PS}^{(1)}$&$r_0m_{\rm PS}^{(2)}$&$r_0m_{\rm V}^{(1)}$&$r_0m_{\rm V}^{(2)}$&$r_0m_{\rm PS}^{(1)}$&$r_0m_{\rm PS}^{(2)}$\\
\hline&&&&\\[-4mm]
6.0   &4.295(28)&4.970(33)&4.644(32)&5.255(36)&4.293(27)&4.971(31)\\
6.1   &4.273(29)&4.987(34)&4.618(33)&5.271(37)&4.277(29)&4.993(33)\\
6.2   &4.286(31)&5.013(37)&4.651(37)&5.316(41)&4.282(31)&5.012(36)\\
6.45  &4.282(36)&5.026(42)&4.655(42)&5.332(46)&4.261(35)&5.001(41)\\
6.7859&3.738(37)&5.182(51)&4.139(46)&5.462(55)&4.825(46)&5.181(50)\\
\hline\hline
\end{tabular}
\caption{Results for the pseudo-scalar and vector meson masses for the data 
sets A and B. Superscripts ``(1)'' and ``(2)'' refer to hopping parameters 
$\kappa_{{\rm c}_1}$ and $\kappa_{{\rm c}_2}$ (see text).}
\label{tab:sim_masses}
\end{center}
\end{table}

\begin{table}
\begin{center}
%\footnotesize
\begin{tabular}{l@{\hspace{1mm}}|c@{\hspace{1mm}}c@{\hspace{1mm}}c@{\hspace{1mm}}c@{\hspace{1mm}}c@{\hspace{1mm}}c@{\hspace{1mm}}c@{\hspace{1mm}}c@{\hspace{1mm}}cc}
\hline\hline\\[-4mm]
& A & A+B & A & A+B & A+B & A+B \\
$\beta$&$r_0(m_{\dsubstar}-m_{\dsub})$&$r_0\fDs$&$r_0\fDsstar/r_0\fDs$&$r_0M_{\rm c}^{({\rm cs})}$&$r_0M_{\rm c}^{({\rm cc})}$&$r_0M_{\rm c}^{({\rm c})}$\\
\hline\\[-3mm]
6.0	&0.283(11)&0.5165(62)&1.018(42)&4.369(47)&5.203(56)&3.236(35)\\
6.1	&0.284(8) &0.5604(67)&1.014(39)&4.272(46)&4.803(52)&3.477(38)\\
6.2	&0.305(14)&0.5809(84)&1.131(68)&4.243(47)&4.640(51)&3.689(41)\\
6.45	&0.310(13)&0.5813(76)&1.142(49)&4.170(47)&4.368(50)&3.931(45)\\ 
6.7859	&0.297(12)&0.5613(81)&1.064(50)&4.070(48)&4.171(50)&3.980(48)\\
\hline\hline\\[-4mm]
c.l.	&0.298(14)&0.557(11)&1.066(61)&4.040(56)&4.055(58)&4.090(54)\\
\hline\hline
\end{tabular}
\caption{Results for the mass splitting, decay constants and the RGI charm
quark mass at finite lattice spacing and in the continuum limit 
(c.l., bottom row).}\label{tab:res}
\end{center}
\end{table}

We analyzed our data using the $\Gamma$-method \cite{Wolff:2003sm}, where the
statistical errors of the observables are estimated by directly analyzing 
autocorrelation functions, and cross-checked its outcome by a jackknife 
procedure.  
Generically, autocorrelation times turned out to be small so that our 
measurements on gauge field ensembles at given $\beta$ could be treated as 
statistically independent.

The results for the pseudo-scalar and vector meson masses are summarized in 
table~\ref{tab:sim_masses}.
For both of the data sets A and B we have computed results for pseudo-scalar
masses very close to the physical point
$r_0m_{\rm D_s}=0.5\,{\rm fm}\times 1969\,{\rm MeV}\approx4.988$ \cite{PDBook}. 
In order to make predictions at the physical point, we interpolate the
mass splitting and the RGI charm quark mass linearly in the pseudo-scalar 
meson mass and the decay constant linearly in the inverse pseudo-scalar 
meson mass, 
as suggested by HQET. 
In the pseudo-scalar channel we have results for primary observables
from the (statistically independent) data sets A and B; 
therefore, we average the results and add the statistical errors in 
quadrature (cf. table \ref{tab:res}).

\begin{figure}
\centering
\psfrag{aaaaaaaaaaa}[l][l][1][0]{\small\hspace{-2mm}this work $\ca$ by ALPHA
	\cite{Luscher:1997ug}}
\psfrag{bbbbbbbbbbb}[l][l][1][0]{\small\hspace{-2mm}Ali-Khan \textit{et al.} \cite{AliKhan:2007tm}}
\psfrag{ccccccccccc}[l][l][1][0]{\small\hspace{-2mm}this work $\ca$ by LANL 
	\cite{Bhattacharya:2001ks}}
\psfrag{ddddddddddd}[l][l][1][0]{\small\hspace{-2mm}UKQCD  
	\cite{Bowler:2000xw}}
\psfrag{xlabel}[t][c][1][0]{$(a/r_0)^2$}
\psfrag{ylabel}[b][b][1][0]{$r_0\fDs$}
\psfrag{aor0sq}[t][c][1][0]{\large$(a/r_0)^2$}
\psfrag{fDs}[b][c][1][0]{\large$r_0 \rm \fDs$}
\psfrag{val}[c][c][1][0]{\large$r_0\rm \fDs=$}
\epsfig{scale=.3,angle=-90,file=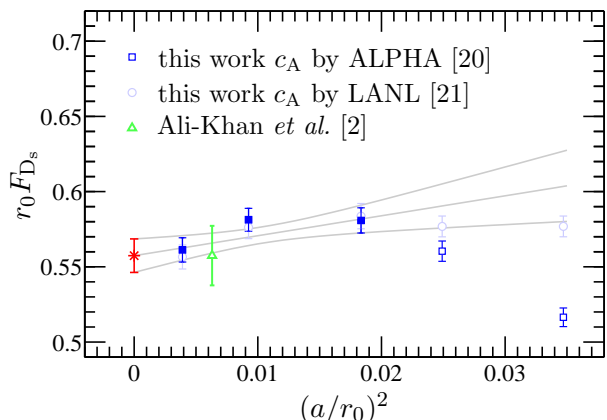}
\caption{Extrapolation to the continuum limit of the decay constant 
${\fDs}$. We also show the results of the recent simulation with 
non-perturbatively improved Wilson fermions in \cite{AliKhan:2007tm}. 
Only the data represented by filled squares entered the fit.}
\label{fig:fDscont}
\end{figure}

\begin{figure}
\centering
\psfrag{aaaaaaaaaaa}[l][l][1][0]{\small\hspace{-2mm}$\ca$ by ALPHA
	\cite{Luscher:1997ug}}
\psfrag{ccccccccccc}[l][l][1][0]{\small\hspace{-2mm}$\ca$ by LANL 
	\cite{Bhattacharya:2001ks}}
\psfrag{xlabel}[t][c][1][0]{$(a/r_0)^2$}
\psfrag{aor0sq}[t][c][1][0]{\large $(a/r_0)^2$}
\psfrag{ylabel}[b][c][1][0]{\large$\fDsstar/\fDs $}
\psfrag{valstar}[c][c][1][0]{\large$\fDs/ \fDsstar=$}
\epsfig{scale=.3,angle=-90,file=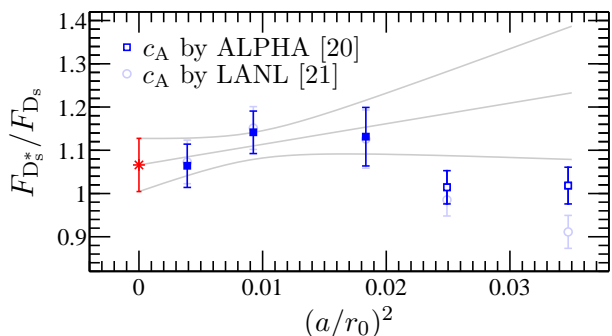}
\caption{Extrapolation to the continuum limit of the ratio ${\fDsstar}/\fDs$. 
Only the data represented by filled squares entered the fit.
Note that this quantity was extracted from data set A only, as the vector
channel correlators are available only in this case.}
\label{fig:fDsratiocont}
\end{figure}

\begin{figure}
\centering
\psfrag{xlabel}[t][c][1][0]{$(a/r_0)^2$}
\psfrag{ylabel}[b][b][1][0]{$r_0(m_{\dsub^\ast}-m_\dsub)$}
\psfrag{aor0sq}[t][c][1][0]{\large $(a/r_0)^2$}
\psfrag{r0mV}[b][c][1][0]{\large$r_0(m_{\rm D_s^\ast}-m_{\rm D_s})$}
\psfrag{val}[c][c][1][0]{$r_0(m_{\rm D_s^\ast}-m_{\rm D_s})=$}
\epsfig{scale=.3,angle=-90,file=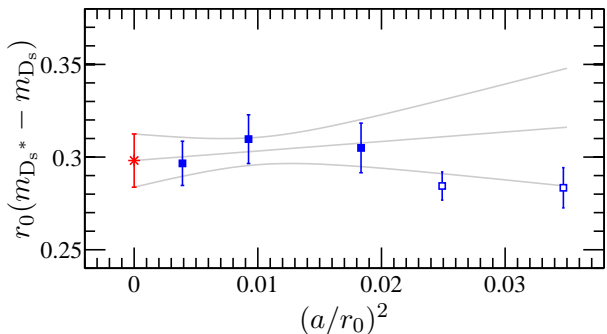}
\caption{Extrapolation to the continuum limit of the mass splitting. 
Only the data represented by filled squares entered the fit.
As in figure~\ref{fig:fDsratiocont}, this quantity was extracted from data 
set A only.}
\label{fig:clms1}
\end{figure}

\begin{figure}
\centering
\psfrag{aaaaaaaa}[lc][l][1][0]{\begin{minipage}{\linewidth}\mbox{}\\[-2.5mm]\hspace*{-2mm}\footnotesize$M_{\rm c}^{\rm (cc)}$\end{minipage}}
\psfrag{b}[lc][l][1][0]{\begin{minipage}{\linewidth}\mbox{}\\[-2.0mm]\hspace*{-2mm}\footnotesize$M_{\rm c}^{\rm (cs)}$\end{minipage}}
\psfrag{c}[lc][l][1][0]{\begin{minipage}{\linewidth}\mbox{}\\[-2.5mm]\hspace*{-2mm}\footnotesize$M_{\rm c}^{\rm (c)}$\end{minipage}}
\psfrag{xlabel}[t][c][1][0]{$(a/r_0)^2$}
\psfrag{ylabel}[b][b][1][0]{$r_0M_{\rm c}$}
\epsfig{scale=.3,angle=-90,file=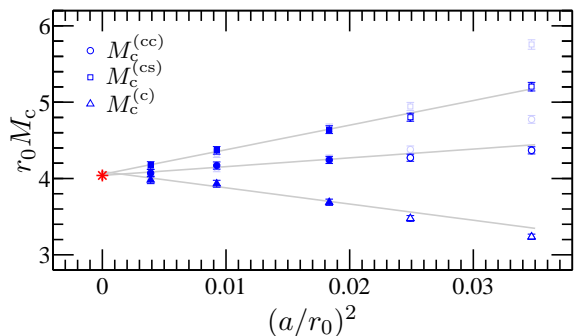}
\caption{Extrapolation to the continuum limit of the RGI mass of the charm
quark. Only the data represented 
by filled symbols entered the fit. For the red star and squares, 
the $\ca$ of \cite{Luscher:1997ug} (ALPHA) 
was used, while it is the one of \cite{Bhattacharya:2001ks} (LANL) 
for the grey data points.}
\label{fig:Mc}
\end{figure}
%
%%%%%%%%%%%%%%%%%%%%%%%%%%%%%%%%%%%%%%%%%%%%%%%%%%%%%%%%%%%%%%%%%%%%%%%%%%%%%%
\section{Discussion}
%%%%%%%%%%%%%%%%%%%%%%%%%%%%%%%%%%%%%%%%%%%%%%%%%%%%%%%%%%%%%%%%%%%%%%%%%%%%%%
The approach of the data to the continuum limit for $r_0\fDs$, 
$\fDsstar/\fDs$, $r_0(m_{\rm D_s^\ast}-m_{\rm D_s})$ and the three definitions 
of the RGI charm quark mass is depicted in 
figures \ref{fig:fDscont}\,--\,\ref{fig:Mc} for all five lattice spacings.
As far as they enter, both sets of non-perturbative values for the 
improvement coefficient $\ca(g_0)$ of the axial-vector current,
i.e.~from ref.~\cite{Luscher:1997ug} (ALPHA) and from
ref.~\cite{Bhattacharya:2001ks} (LANL) and differing in the particular
improvement conditions imposed for their determination, have been 
considered.

In ref.~\cite{Kurth:2001yr} 
a breakdown of ${\rm O}(a)$ improvement for improved Wilson quarks with
heavy quark masses about $aM_{\rm h}\gtrsim 0.64$ was found in 
perturbation theory.
Since the charm
quark mass in lattice units is heavier than this bound for the two coarsest 
lattices with $a= 0.093\,$fm and $a=0.079\,$fm, we exclude the results on
these lattices from the continuum extrapolations. 
In fact, 
significant deviations from the linear dependence on $a^2$ of the data
on the coarser lattices is observed for $r_0\fDs$ and $\fDsstar/\fDs$,
and also for the charm quark mass definitions $r_0M_{\rm c}^{(\rm cs)}$ and
$r_0M_{\rm c}^{(\rm cc)}$; in the latter case, however, only 
when using $\ca$ by LANL. 
In general, on the coarsest two lattices we see significant discrepancies 
between the results of our non-perturbative computation obtained with the 
ALPHA-$\ca$ and the LANL-$\ca$, respectively, which is in line with 
perturbative criterion on $aM_{\rm h}$ consulted above. 

For the final results in the continuum limit of quenched QCD we quote the
linear extrapolation of the data at the three finest lattice spacings, 
where we employ the $\ca$ of \cite{Luscher:1997ug} for the improvement of 
the axial current.
In this way we are well within the range of quark masses $aM_{\rm h}$,
for which ${\rm O}(a)$ improvement and thus the validity of the 
$a^2$-expansion and the associated scaling behaviour are expected to hold.
In addition, the ambiguity owing to the choice of $\ca$ we observe has no 
significant effect at these lattice spacings.
 These data points are fully compatible with
$(a\Lambda_{\rm QCD})^2$ corrections to the continuum limit when
using $\Lambda_{\rm QCD}\approx 500$ MeV which also indicates that there
are no large $(aM_c)^2$-corrections.

The fact that the continuum limit of the 
three definitions of the RGI charm quark mass 
(\ref{eqn:Mhs})\,--\,(\ref{eqn:Mh})
yield compatible continuum extrapolated results 
demonstrates nicely the universality of the continuum limit and thereby 
provides strong evidence for the correctness of the extrapolation procedure.
Our final results in units of the hadronic scale $r_0$ are 
collected in the bottom row of table~\ref{tab:res}. 
The quoted errors include all errors, i.e.~the statistical ones,
those stemming from the fits at intermediate stages of the analysis as well 
as the uncertainties of $r_0/a$ and the various $Z$-factors. 
Using $r_0=0.5\,$fm, we also convert our numbers to physical units: 
$\fDs=220(4)(?)\,$MeV, $m_{\rm D_s^\ast}-m_{\rm D_s}=118(6)(?)\,$MeV, 
$M_{\rm c}^{({\rm cs})}=1.59(2)(?)\,$GeV,
$M_{\rm c}^{({\rm cc})}=1.60(2)(?)\,$GeV and
$M_{\rm c}^{({\rm c})}=1.61(2)(?)\,$GeV.
With the second error we indicate the unknown systematic contribution due 
to the quenched approximation, by which also the result for the 
dimensionless ratio $\fDsstar/\fDs$ is affected.
Note that thanks to the additional, finer lattice spacing and the 
substantially increased total statistics in the pseudo-scalar channel,
we have gained a factor of about two in accuracy for $r_0M_{\rm c}$ and
$r_0\fDs$ compared to the earlier investigations 
of~\cite{Rolf:2002gu,Juttner:2003ns}.

Whereas in the case of the RGI charm quark mass we observe agreement within 
errors with the result of ref.~\cite{Rolf:2002gu}, 
the continuum limit of $r_0\fDs$ quoted 
in~\cite{Juttner:2003ns},
which is based on data set A for the range of lattice spacings 
$a\approx0.08-0.05\,$fm, 
now proves to overestimate our result in
table~\ref{tab:res} for the leptonic decay constant of the $\dsub$-meson 
by about three standard deviations.
We have identified two reasons for this: Firstly, after increasing the 
statistics for $r_0\fDs$ 
by generating data set B, the central value at $a=0.048\,$fm ($\beta=6.45$)
moved down by about $1.5\sigma$. 
Secondly, in addition to \cite{Juttner:2003ns} we now have results for a  
finer lattice spacing with $a= 0.031\,$fm ($\beta=6.7859$).
Thus, increasing the statistics by
combining data set A and B, discarding the data point at $a=0.079\,$fm 
($\beta=6.2$) from the continuum extrapolation and including the results 
at $a= 0.031\,$fm instead causes a shift in the central value
for $\fDs$ by about three standard deviations compared to the result in 
\cite{Juttner:2003ns}.
Statistical fluctuations in conjunction with leading plus possibly
higher-order lattice artefacts can misguide the true continuum extrapolation.

This clearly reveals that it is indispensable to incorporate as small
lattice spacings as possible into computations in the charm sector of  
${\rm O}(a)$ improved lattice QCD, in order to get a controlled handle on
the continuum limit.
%%%%%%%%%%%%%%%%%%%%%%%%%%%%%%%%%%%%%%%%%%%%%%%%%%%%%%%%%%%%%%%%%%%%%%%%%%%%%%
\section{Conclusion}
%%%%%%%%%%%%%%%%%%%%%%%%%%%%%%%%%%%%%%%%%%%%%%%%%%%%%%%%%%%%%%%%%%%%%%%%%%%%%%
%
Full QCD simulations with improved Wilson fermions in large volumes and
at small lattice resolutions very close to the physical
point are now feasible \cite{DelDebbio:2006cn,DelDebbio:2007pz}. 
Since their formulation is free of conceptual problems and since they are 
comparably cheap to simulate, they are now being applied on a large scale 
in the computation of Standard Model parameters at very high precision. 
The results of this work, although referring to the quenched approximation, 
emphasize the importance of performing a scaling study down to very small 
lattice spacings when calculating observables for hadrons containing a 
charm quark. 
Without the data at the finest lattice spacing and its satisfactory 
statistical accuracy obtained here, the continuum extrapolation would have 
yielded too large a value for $\fDs$ and also for $\fDsstar/\fDs$.

It is expected that these findings for the quenched theory carry over to 
the theory with dynamical sea quarks. We stress that only a scaling study of 
charmed observables down to extremely fine lattice spacings will allow for
a reliable assessment of the cutoff effects. These findings, however, do not
directly apply to other fermion discretizations and therefore, similar 
studies are crucial in each individual case.
\acknowledgments
We are grateful to M.~Della Morte, Stefan Schaefer,
R.~Sommer and H.~Wittig for fruitful discussions and a 
critical reading of the manuscript.
We would like to thank P.~Fritzsch for his contributions in adapting and
optimizing the code for the computation of the Schr\"odinger functional 
correlation functions on apeNEXT.
This work was in part based on the MILC Collaboration's public lattice 
gauge theory code, see~\cite{MILC}.
We thank NIC for allocating computer time on the APE computers at DESY
Zeuthen to this project and the APE group for its help.
We further acknowledge partial support by the Deutsche 
Forschungsgemeinschaft (DFG) under grant HE 4517/2-1 as well as by the 
European Community through EU Contract No.~MRTN-CT-2006-035482, 
``FLAVIAnet''.

%%%%%%%%%%%%%%%%%%%%%%%%%%%%%%%%%%%%%%%%%%%%%%%%%%%%%%%%%%%%%%%%%%%%%%%%%%%%%%
%
\bibliographystyle{JHEP}
\bibliography{fds2}
\end{document}